\documentclass[aps,pra,10pt,reprint,superscriptaddress]{revtex4-2}
\usepackage{amsmath,amsfonts,amssymb}
\usepackage{xcolor,graphicx}
\usepackage{braket}
\usepackage{bm}
\usepackage{overpic}
\usepackage{float}
\usepackage[pdftex]{hyperref}
\usepackage{csquotes}
\hypersetup{
	colorlinks = true,
	linkcolor = blue,
	anchorcolor = blue,
	citecolor = red,
	filecolor = blue,
	urlcolor = blue}

\begin{document}

\title{Impact of atomic initial conditions on nonclassicality of the light in the ladder-type three-level Jaynes-Cummings model}

\author{L. Hernández-Sánchez}
\email[e-mail: ]{leonardi1469@gmail.com}
\affiliation{Centro de Bachillerato Tecnológico Industrial y de Servicios No. 144: José Emilio Grajales Moguel, Boulevard Mayor Sabines No. 1982, Col. 24 de junio, 29047 Tuxtla Guti\'errez, Chiapas, Mexico}
\affiliation{Instituto Nacional de Astrofísica Óptica y Electrónica, Calle Luis Enrique Erro No. 1\\ Santa María Tonantzintla, Puebla, 72840, Mexico}
\author{A. Flores-Rosas}
\affiliation{Facultad de Ciencias en F\'isica y Matem\'aticas, Universidad Aut\'onoma de Chiapas, Carretera Emiliano Zapata, Km. 8, Rancho San Francisco, 29050 Tuxtla Guti\'errez, Chiapas, Mexico}
\author{S. Mendoza Vásquez}
\affiliation{Facultad de Ciencias en F\'isica y Matem\'aticas, Universidad Aut\'onoma de Chiapas, Carretera Emiliano Zapata, Km. 8, Rancho San Francisco, 29050 Tuxtla Guti\'errez, Chiapas, Mexico}
\author{I. Ramos-Prieto}
\affiliation{Instituto Nacional de Astrofísica Óptica y Electrónica, Calle Luis Enrique Erro No. 1\\ Santa María Tonantzintla, Puebla, 72840, Mexico}
\author{F. Soto Eguibar}
\affiliation{Instituto Nacional de Astrofísica Óptica y Electrónica, Calle Luis Enrique Erro No. 1\\ Santa María Tonantzintla, Puebla, 72840, Mexico}
\author{H. M. Moya-Cessa}
\affiliation{Instituto Nacional de Astrofísica Óptica y Electrónica, Calle Luis Enrique Erro No. 1\\ Santa María Tonantzintla, Puebla, 72840, Mexico}

\date{\today}

\begin{abstract}
We explore the interaction between a three-level atom and a single-mode quantized cavity, known as the three-level ladder-type Jaynes-Cummings model. By employing the exact solution of the Schrödinger equation, we investigate how the initial conditions of the atom influence the occupation probabilities of the atomic energy levels, average photon number, and the nonclassicality of light, assessed through the Mandel $\mathcal{Q} (t)$ parameter and the Wigner function. Our findings are rigorously validated through comprehensive numerical simulations, ensuring robust and consistent outcomes.
\end{abstract}
\maketitle

\section{Introduction}\label{Introducción}
The Jaynes-Cummings model (JCM) has been widely recognized and studied as a fundamental cornerstone in the field of quantum optics. This essential theoretical framework provides a detailed description of the interaction between a two-level atom and a single-mode electromagnetic field in a lossless cavity~\cite{JC_1963}. Constructed under the dipole and rotating-wave approximations, this model is considered the most fundamental for studying the interaction between matter and field in the realm of quantum optics, owing to its exactly integrable solutions, making it a powerful tool for exploring quantum dynamics without resorting to perturbative approximations~\cite{Stemholm_1973,Bruce_1993,Larson_2022}.

Over the years, numerous extensions and generalizations of the original JCM have been investigated to address more complex and realistic aspects of the radiation-matter interaction. The mathematical description of these models becomes more intricate, making solving the Schrödinger equation associated with them an even more challenging task~\cite{Gerry_Book, Klimov_2009, Agarwal_2012, Meystre_2021}. These extensions include atoms with multiple energy levels~\cite{Abdel_1987, Buzek_1990}, multiple modes of the electromagnetic field~\cite{Sukumar_1981, Abdalla_1991}, losses~\cite{Saavedra_1996, Scala_2007}, interaction with external fields~\cite{Alsing_1992, Dutra_1994, Bocanegra_2023, Hernandez_book}, to name just a few.

Furthermore, the JCM exhibits significant non-classical properties and sub-Poissonian and super-Poissonian photon statistics under different conditions of the electromagnetic field cavity~\cite{Gerry_Book, Meystre_2021}. Furthermore, it has been shown that atomic transitions between the two levels of the atom differ depending on whether the atom is initially in its excited state or its ground state~\cite{Hernandez_2023_a, Hernandez_book}. Notably, the collapses and revivals of atomic transitions have already been experimentally observed~\cite{Eberly_1980, Meschede_1985, Haroche_2006}.

However, it has been demonstrated that three-level atoms offer a more complex framework for understanding and analyzing quantum phenomena compared to two-level atoms, thereby allowing for the exploration of a wider variety of effects and physical processes~\cite{Scully_1997, Haroche_2006}. The multifaceted nature of three-level atoms provides an ideal platform for investigating phenomena such as quantum coherence, decoherence~\cite{Cardimona_1989, Nath_2003}, and the effects of interaction with external electromagnetic fields~\cite{Kumar_2023}. From an experimental standpoint, the study of three-level atoms is essential for understanding and developing technologies based on quantum systems, such as quantum devices and quantum information systems. For instance, in the context of the maser (microwave amplification by stimulated emission of radiation), the three-level atom plays a crucial role in signal amplification, enabling the generation of coherent and high-intensity microwaves~\cite{Yoo_1985, Haroche_2006}.

In this study, we are interested in investigating the nonclassical properties of light in the JCM associated with a three-level atom interacting in a lossless single-mode cavity, known as the ladder-type three-level JCM. Specifically, we analyze how different initial conditions in the atom influence the system dynamics and its statistical behavior, while the cavity is initially prepared in a coherent state. The structure of this work is as follows: in Section~\ref{Model}, we apply a time-dependent unitary transformation and, using the traditional method to solve the dynamics of the JCM, we find the exact solution of the Schrödinger equation associated with the Hamiltonian that models the system. Then, in Section~\ref{Dynamics}, we analyze how different initial conditions of the atom influence its occupation probabilities across different energy levels, as well as the variation in the expected photon number. In Section~\ref{Nonclassical}, we explore the nonclassical properties of light associated with this model through the study of the Mandel $\mathcal{Q}(t)$ parameter  and the Wigner distribution. Finally, in Section~\ref{Conclusiones}, we present our conclusions.

\section{The ladder-type three-level model}\label{Model}
\begin{figure}[H]
\centering
\begin{overpic}[width=\linewidth]{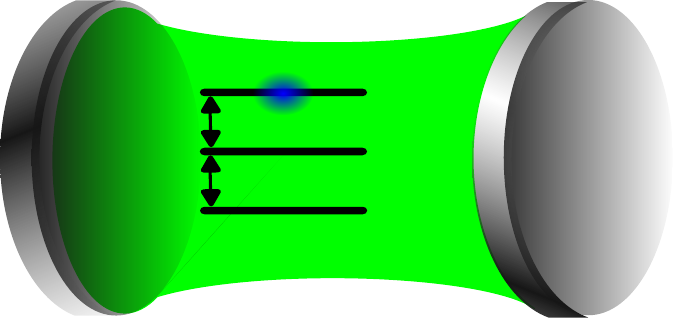}
        \put(53.2,35){ $\ket{n,3}$}
        \put(53.2,25.8){ $\ket{n+1,2}$}
        \put(53.2,17.5){ $\ket{n+2,1}$}
        \put(23.5,31.5){ \large $\omega_0$}
        \put(23.5,22.5){ \large $\omega_0$}
        \put(65,10){ \large $\omega_c$}
\end{overpic}
\caption{Scheme of a lossless cavity formed by perfectly reflecting mirrors. Inside the region bounded by these mirrors, a three-level atom, characterized by a transition frequency $\omega_{0}$, interacts with a cavity field having a frequency $\omega_c$.}
\label{fig1}
\end{figure}
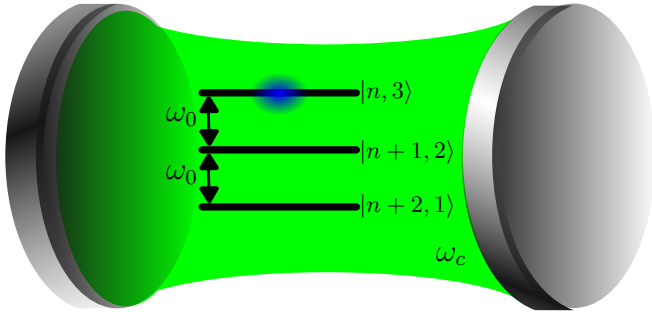

Let us consider an atom with three energy levels: $\ket{1}$ representing the lowest energy level, $\ket{2}$ the intermediate level, and $\ket{3}$ the highest level. The transition frequency between these levels is constant and denoted as $\omega_0$. This atom is situated within a cavity formed by perfectly reflecting mirrors, which maintain a single quantized mode of electromagnetic field with a frequency of $\omega_c$. This configuration is illustrated in Fig.~\ref{fig1}.

From Fig.~\ref{fig1}, we can observe that this system, known in the literature as a ladder-type three-level system ($\varXi$), is characterized by allowing only transitions of the form $\ket{1} \, \leftrightarrow \, \ket{2} \, \leftrightarrow \, \ket{3}$. To describe these transitions mathematically, we employ the special unitary group SU(2); this $8$-dimensional Lie group consists of unitary $3 \times 3$ matrices with determinant $1$, known as \textit{Gell-Mann matrices}. These matrices generalize the Pauli matrices used for two-level systems. For this particular model, the atomic part of the system can be described by the operators $\hat{I}_{+} = \ket{3}\bra{2} + \ket{2}\bra{1}$, $\hat{I}_{-} = \ket{2}\bra{3} + \ket{1}\bra{2}$, and $\hat{I}_{z} = \ket{3}\bra{3} - \ket{1}\bra{1}$, which satisfy the commutation relations $[\hat{I}_{+}, \hat{I}_{-}] = \hat{I}_{z}$ and $[\hat{I}_{z} , \hat{I}_{\pm}] = \pm \hat{I}_{\pm}$. From these matrices, the Hamiltonian describing the system can be written as~\cite{Nath_2003, Hernandez_2017}
\begin{equation}\label{Hamiltoniano_1}
\hat{H}  = \omega_{0} \hat{I}_{z} +  \omega_c\hat{a}^{\dagger} \hat{a} +  g \left( \hat{I}_{+}\hat{a} + \hat{I}_{-} \hat{a}^{\dagger} \right),
\end{equation}
where $g$ is the coupling constant between the three-level system and the cavity field under the dipolar approximation. We assume $g$ is consistent across all levels. As usual, the creation and annihilation operators, $\hat{a}^{\dagger}$ and $\hat{a}$, describe the cavity field mode, satisfying the bosonic commutation relation $[\hat{a}, \hat{a}^{\dagger}] = 1$. Additionally, for convenience, we have adopted the convention of setting $\hbar = 1$ (reduced Planck constant) throughout this study.

To tackle the Schrödinger equation for this system, we employ a time-dependent unitary transformation $\hat{\mathcal{
T}}=\exp\left[\mathrm{i} \omega_c t (\hat{n}+ \hat{I}_z)\right]$, leading to the interaction representation, where the Hamiltonian is given by
\begin{equation}\label{Hamiltoniano_2}
\begin{split}
\hat{\mathcal{H}} & = \hat{\mathcal{T}}\hat{H} \hat{\mathcal{T}}^{\dagger}-\mathrm{i}\hat{\mathcal{T}}\partial_t\hat{\mathcal{T}}^\dagger,\\
& = \Delta \hat{I}_z + g \left( \hat{I}_{+}\hat{a} + \hat{I}_{-} \hat{a}^{\dagger} \right),
\end{split}
\end{equation}
with $\Delta = \omega_{0} - \omega_c$ representing the detuning between the unimodal field frequency and the atomic transition frequency.

To solve the Schrödinger equation in the interaction picture, we follow the traditional approach of expanding the atom-field state vector at time $t$ as a linear combination or superposition of Fock states $\left\{ \ket{n} \right\}$~\cite{Nath_2003, Gerry_Book, Hernandez_2017}. Since the model only allows atomic transitions of the form $\ket{1} \, \leftrightarrow \, \ket{2} \, \leftrightarrow \, \ket{3}$, this superposition can be written as
\begin{equation}\label{Psi}
\begin{split}
\ket{ \Psi (t) } & =\sum_{n=0}^{\infty}\left[C_3 (t) \ket{n,3} + C_2(t) \ket{n+1,2} \right. \\
& \hspace{1.5cm} + \left. C_1(t) \ket{n+2,1} \right],
\end{split}
\end{equation}
which reduces the problem to solving the following system of coupled ordinary differential equations
\begin{align}
\mathrm{i}\frac{d}{dt}
    \begin{bmatrix}
        C_3(t)\\C_2(t)\\C_1(t)
    \end{bmatrix}
    =& 
    \begin{bmatrix}
        \Delta & g\sqrt{n+1} & 0 \\
        g\sqrt{n+1} & 0 & g\sqrt{n+2} \\
        0 & g\sqrt{n+2} & -\Delta
    \end{bmatrix}
    \begin{bmatrix}
        C_3(t)\\C_2(t)\\C_1(t)
    \end{bmatrix}.
\end{align}

The general solution to these differential equations can be quite laborious, but it simplifies significantly when we consider that the atomic transition frequencies and the cavity field frequency are in resonance, i.e., when $\Delta=0$. In this case, the solution can be expressed as 
\begin{align}\label{Sistema}
    \begin{bmatrix}
        C_3(t)\\C_2(t)\\C_1(t)
    \end{bmatrix}
    = 
    \begin{bmatrix}
        M_{11}(t) & M_{12}(t) & M_{13}(t)\\
        M_{21}(t) & M_{22}(t) & M_{23}(t)\\
        M_{31}(t) & M_{32}(t) & M_{33}(t)
    \end{bmatrix}
    \begin{bmatrix}
        C_3(0)\\C_2(0)\\C_1(0)
    \end{bmatrix},
\end{align}
where the quantities $|C_3(0)|^2$, $|C_2(0)|^2$, and $|C_1(0)|^2$ determine the initial distribution of photons in the field at the upper, intermediate, and lower levels of the atom, respectively. Meanwhile, $\beta_n = g \sqrt{2n+3}$ represents the generalized Rabi frequency. Additionally, the functions $M_{ij}(t)$, with $i,j = 1,2,3$, are determined by
\begin{equation}\label{C_nD_n}
\begin{split}
M_{11}(t) & = \frac{g^2}{\beta_{n}^2} \left[  (n+1) \cos(\beta_{n} t) + (n+2)  \right], \\ 
M_{12}(t) & = -\mathrm{i} \frac{g\sqrt{n+1}}{\beta_{n}} \sin(\beta_{n} t), \\
M_{13}(t) & = \frac{g^2 \sqrt{(n+1)(n+2)}}{\beta_{n}^2} \left[ \cos(\beta_{n} t) - 1 \right], \\
M_{21}(t) & = -\mathrm{i} \frac{g \sqrt{n+1}}{\beta_{n}} \sin(\beta_{n} t), \\
M_{22}(t) & = \cos(\beta_{n} t), \\
M_{23}(t) & = -\mathrm{i} \frac{g \sqrt{n+2}}{\beta_{n}} \sin(\beta_{n} t), \\
M_{31}(t) & = \frac{g^2 \sqrt{(n+1)(n+2)}}{\beta_{n}^2} \left[ \cos(\beta_{n} t) - 1 \right], \\
M_{32}(t) & = -\mathrm{i} \frac{g \sqrt{n+2}}{\beta_{n}} \sin(\beta_{n} t), \\
M_{33}(t) & = \frac{g^2 }{\beta_{n}^2} \left[(n+2) \cos(\beta_{n} t) + (n+1)  \right].
\end{split}
\end{equation}

\begin{table*}[ht]
\centering
\caption{Occupation probabilities of the atomic energy levels}
\begin{tabular}{|c|c|c|c|}
\hline
& Atom initially in the upper state & Atom initially in the intermediate state & Atom initially in the lower state \\
\hline
$P^{3}(t) =$ & $\sum_{n=0}^{\infty} |C_3 (0)|^2 |M_{11} (t)|^2$ & $\sum_{n=0}^{\infty} |C_2 (0)|^2 |M_{12} (t)|^2$ & $\sum_{n=0}^{\infty} |C_1 (0)|^2 |M_{13} (t)|^2$ \\
\hline
$P^{2}(t) = $ & $\sum_{n=0}^{\infty} |C_3 (0)|^2 |M_{21} (t)|^2$ & $\sum_{n=0}^{\infty} |C_2 (0)|^2 |M_{22} (t)|^2$ & $\sum_{n=0}^{\infty} |C_1 (0)|^2 |M_{23} (t)|^2$ \\
\hline
$P^{1}(t) = $ & $\sum_{n=0}^{\infty} |C_3 (0)|^2 |M_{31} (t)|^2$ & $\sum_{n=0}^{\infty} |C_2 (0)|^2 |M_{32} (t)|^2$ & $\sum_{n=0}^{\infty} |C_1 (0)|^2 |M_{33} (t)|^2$ \\
\hline
\end{tabular}
\label{table}
\end{table*}

\section{Dynamics}\label{Dynamics}
Leveraging the results obtained from Eq.~\eqref{Sistema}, the solution to the Schrödinger equation proposed in Eq.~\eqref{Psi} is fully determined. This solution empowers us to calculate and analyze any observable or dynamical variable of the system, as will be demonstrated subsequently. The only requirement is to specify the initial conditions for both the atom and the cavity field, represented by $\ket{\Psi (0)} = \ket{\Psi^F (0)} \otimes \ket{\Psi^A (0)} $. For simplicity, we assume that the cavity field is initially in a coherent state  $\ket{\alpha}$, where $\alpha$ is an arbitrary complex number, while the atom can reside in any of its three energy levels.

\subsection{Atomic occupation probabilities}
In scientific literature, the study of the interaction between the atom and the cavity field often emphasizes the occupation probabilities of the energy levels of the atom. This focus is crucial for understanding the dynamics of the system, as occupation probability indicates the number of atoms present in a specific energy state at any given time~\cite{Nath_2003,Gerry_Book}.

Within the framework of the analyzed model, the atom can exist in one of three distinct energy levels: the lower state $(\ket{1})$, the intermediate state $(\ket{2})$, and the upper state $(\ket{3})$. The occupation probabilities for each level can be determined by calculating the expected value of the projection operators corresponding to each initial atomic condition. These results are summarized in Table~\ref{table}.

Here, $P^{j}(t) = |C_j (t) |^2$ for $j=1,2,3$. Additionally, $\left|C_n(0)\right|^2 = P_n$,  $|C_2(0)|^2 = P_{n+1}$, and $|C_1(0)|^2 = P_{n+2}$, where $P_n$ representing the photon probability distribution associated with the coherent state~\cite{Gerry_Book, Meystre_2021}
\begin{equation} \label{P_n_coherent}
P_{n} = e^{-|\alpha|^2} \frac{|\alpha|^{2n}}{n!}.
\end{equation}

The assignments of $|C_2(0)|^2$ and $|C_1(0)|^2$ have significant physical meaning. A detailed analysis of equation~\eqref{Psi}, which describes the wave function of the complete system, clarifies our initial assumption of one additional energy quantum in the intermediate level and two additional energy quanta in the lower level compared to the upper level. Therefore, if the atom is in the intermediate or ground state, the probability of having zero photons in any of these levels within the field is zero.

\begin{figure*}[!t]
\centering
\includegraphics[width=1\linewidth]{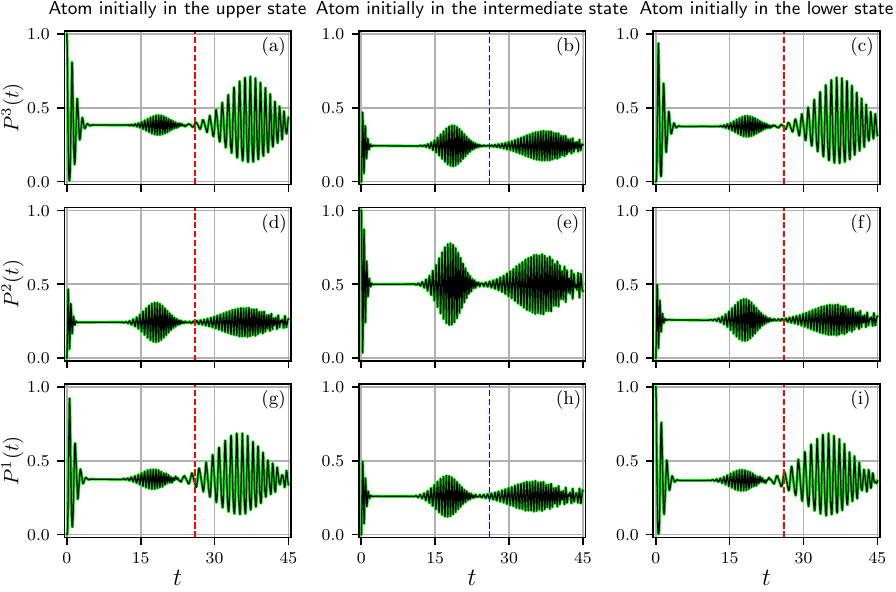}
\caption{Atomic occupation probabilities corresponding to the initial condition of the cavity field prepared in a coherent state, using the following parameter values: $\omega_c = \omega_{eg} = 0.3$, $g = 1.0$, and $\alpha = 4$. Subfigures depict the occupation of the atom in the upper state (left), intermediate state (center), and lower state (right) based on the initial condition of the atom being in the upper, intermediate, or lower state, respectively. The black lines represent the analytical results, while the green lines depict the numerical solutions obtained using QuTiP~\cite{Qutip}.}
\label{fig2}
\end{figure*}

In Fig.~\ref{fig2}, we detail the atomic occupation probabilities when the cavity field is initially prepared in a coherent state, using parameter values $\omega_c = \omega_{eg} = 0.3$, $g = 1.0$, and $\alpha = 4$. The subfigures illustrate the probability of occupation of the atom in the upper state (left), intermediate state (center), and lower state (right) based on its initial condition.

The subfigures corresponding to the initial condition of the atom in the upper and lower states (left and right, respectively) reveal that the atomic occupation probabilities for all three states (upper, intermediate, and lower) exhibit very similar population dynamics [see dashed red lines], owing to the initially large average number of photons. Examining the central subfigure, where the atom starts in the intermediate state, one might expect equivalent probabilities of transitioning to the upper and lower states. However, a closer inspection [see dashed blue lines] shows a slight difference in population dynamics. This distinction primarily arises because the lower state initially contains two more photons than the upper state. This difference becomes more pronounced as the average number of photons decreases and diminishes as this number increases.

\subsection{Average photon number}
Another important observable to analyze is the expectation value of the number operator $\hat{n}$, which indicates how the average number of photons evolves over time. This is crucial because it provides a better understanding of the statistical properties of the system, including the photon distribution and its relation to the dynamics of the atom-field interaction. In the context of the three-level model, the expectation value of the number operator $\hat{n}$ is expressed as
\begin{equation}\label{Average photon}
\begin{split}
\braket{ \hat{n}} & = \braket{ \Psi (t) | \hat{n} | \Psi (t)}\\
& = \overline{n} + \sum_{n=0}^{\infty} \left[ \; |C_2 (t) |^2 + 2|C_1 (t) |^2 \; \right].
\end{split}
\end{equation}
From this result, we observe that the average number of photons over time, $\braket{\hat{n}(t)}$, will remain around the initial average photon number $\overline{n}$, except for transitions related to the second term in \eqref{Average photon}. Specifically, the following results are obtained for the different initial atomic conditions:\\

\textbf{Case I}:  Atom initially in the upper state
\begin{equation} \label{N_3}
\begin{aligned}
    \braket{\hat{n}} = \overline{n} & + \sum_{n=0}^{\infty} |C_3 (0) |^2 \left(\frac{g}{\beta_n}\right)^2 (n+1) \bigg\{ \sin^2 (\beta_n t) \\
    &  \hspace{0.6cm}   + 2 \left(\frac{g}{\beta_n}\right)^2 (n+2)  \left[ \cos (\beta_n t) - 1  \right]^2 \bigg\}.
\end{aligned}
\end{equation}

\textbf{Case II}: Atom initially in the intermediate state
\begin{equation} \label{N_2}
\begin{aligned}
    \braket{\hat{n}} = \overline{n} & +
    \sum_{n=0}^{\infty} |C_1 (0) |^2 \bigg\{ \cos^2 (\beta_n t) + 2 \left(\frac{g}{\beta_n}\right)^2
    \\
    & \hspace{3.0cm}  \times (n+2) \sin^2 (\beta_n t) \bigg\}.
\end{aligned}
\end{equation}

\textbf{Case III}: Atom initially in the lower state
\begin{equation} \label{N_1}
\begin{aligned}
    \braket{\hat{n}} & = \overline{n} - \sum_{n=0}^{\infty} |C_1 (0) |^2 \left(\frac{g}{\beta_n}\right)^2 \bigg\{(n+2) \sin^2 (\beta_n t)  \\
    &  \hspace{0.5cm} + 2 \left(\frac{g}{\beta_n}\right)^2 \left[ (n+2) \cos (\beta_n t) + (n+1) \right]^2 \bigg\}.
\end{aligned}
\end{equation}

\begin{figure}[H]
\centering
\includegraphics[width=1\linewidth]{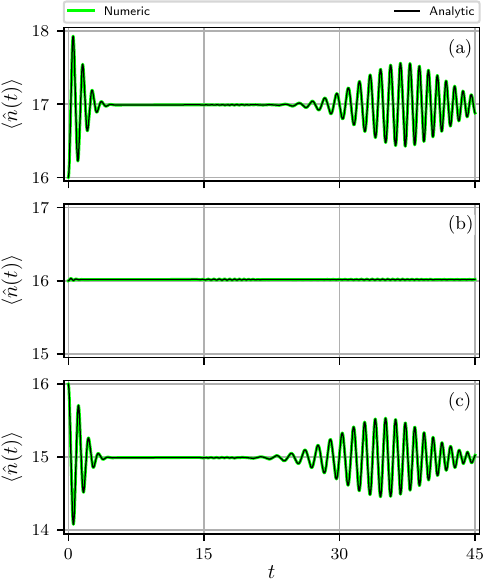}
\caption{The average photon number $\braket{\hat{n}(t)}$ corresponds to the same initial conditions and parameters as those used in Fig.~\ref{fig2}. In (a), the average photon number is shown with the atom initially in the upper level; in (b), with the atom in the intermediate level; and in (c), with the atom in the lower level, respectively. The black lines represent the analytical results, while the green lines depict the numerical results.}
\label{fig3}
\end{figure}

In Fig.~\ref{fig3}, the expected value of the photon number operator $\hat{n}$ is shown for the same parameters as in Fig.~\ref{fig2}, under different initial atomic conditions. Firstly, in Fig.~\ref{fig3}(a), it can be observed that the average number of photons centers around 17, one photon more than its initial state, because the upper level $\ket{3}$ has one more photon than the middle level $\ket{2}$, and two more photons than the lower level ($\ket{1}$). In Fig.~\ref{fig3}(b), we depict the case when the atom is initially in its intermediate state, as per equation \eqref{N_2}. It is evident that due to the absence of transitions between the upper and lower levels, the average number of photons remains unchanged at the initial average photon number ($\overline{n} = |\alpha|^2 = 16$). Finally, in Fig.~\ref{fig3}(c), we illustrate the scenario when the atom starts in its lower state, given by equation \eqref{N_1}; in this case, we observe a very similar behavior to case (a), but now the average centers around 15, one photon less than its initial state, because the lower level $\ket{1}$ has one photon less than the middle level $\ket{2}$, and two photons less than the upper level $\ket{3}$.

\section{Nonclassical properties}\label{Nonclassical}
In this section, we derive criteria to detect the nonclassicality in the considered quantum state, exploring how the atomic initial conditions impact the nonclassicality of light in the ladder-type three-level JCM. This analysis will allow us to better understand how different atomic initial states affect the nonclassical properties of the coupled light field.

\subsection{Mandel $\mathcal{Q} $ parameter}
To analyze the photon statistics of a single-mode radiation field, we consider Mandel's $\mathcal{Q}$ parameter, defined as \cite{Mandel_1979}
\begin{equation}\label{Q}
\mathcal{Q} = \frac{\langle\hat{n}^2\rangle - \langle\hat{n}\rangle^2}{\langle\hat{n}\rangle} - 1.
\end{equation}
When $\mathcal{Q} = 0$, it indicates a Poissonian distribution, while for values $-1 \leq \mathcal{Q} < 0$ ($\mathcal{Q} > 0$), the field exhibits sub-Poissonian (super-Poissonian) photon statistics. Importantly, the negativity of $\mathcal{Q}$ is not a necessary criterion to differentiate quantum states into classical and nonclassical regimes; rather, it serves as a sufficient condition.  There are instances where a state may exhibit nonclassical behavior even when $\mathcal{Q}$ is positive~\cite{Agarwal_2012}.

\begin{figure}[h]
\centering
\includegraphics[width=1\linewidth]{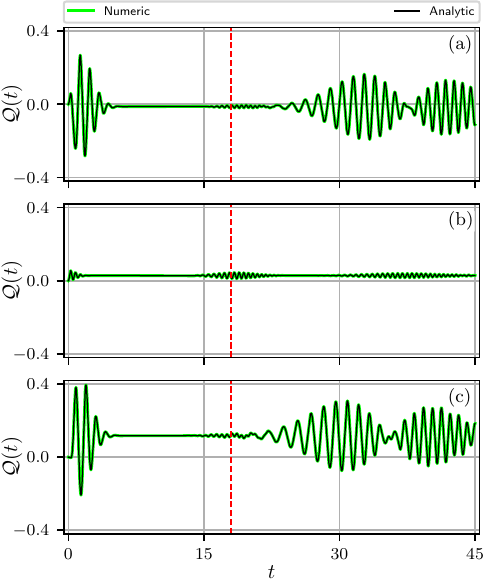}
\caption{The Mandel $\mathcal{Q}(t)$ parameter for the same initial conditions and parameters as those used in Fig.~\ref{fig2}. In (a), the Mandel $\mathcal{Q}(t)$ parameter is shown with the atom initially in the upper level; in (b), with the atom in the intermediate level; and in (c), with the atom in the lower level, respectively. As in previous figures, the black lines depict the analytical results, while the green lines depict the numerical results.}
\label{fig4}
\end{figure}

In Fig.~\ref{fig4}, Mandel's $\mathcal{Q} $ parameter is shown for different atomic initial conditions, with the cavity field initially in a coherent state and using the same parameter values as in the previous figures. From this figure, it is observed that Mandel's $\mathcal{Q}$ parameter exhibits oscillatory behavior with varying amplitudes. The negativity of $\mathcal{Q}$ confirms the nonclassical nature of the considered cavity field state. Specifically, in Fig.~\ref{fig4}, it is evident that when the atom is initially in the upper state, greater nonclassicality is observed compared to cases where the atom starts in the intermediate or lower states. It is important to note that for the intermediate case, the system consistently exhibits classical behavior, albeit closely approaching the nonclassical limit. Finally, when the atom is in its lower level, the behavior is predominantly classical, except for minor negative contributions; this can be attributed to the two additional photons in the lower level compared to the upper level, as nonclassicality decreases with increasing $\alpha$ \cite{Gerry_Book, Kumar_2023}.

\subsection{Wigner distribution}
A way to analyze the nonclassical properties of a state is by examining its distribution in phase space using the Wigner function. This function establishes a direct connection between the density operator of a quantum system and a distribution in phase space, providing a comprehensive representation of the state of the quantum system \cite{Gerry_Book, Agarwal_2012}. The Wigner function can be expressed in series form as follows~\cite{Moya_Book}
\begin{equation}
W(\beta, \beta^*) = \frac{2}{\pi} \sum_{k=0}^{\infty} (-1)^k \braket{ \beta, k| \hat{\rho} | \beta, k},
\end{equation}
where $\ket{\beta, k} = \hat{D}(\beta) \ket{k}$ is the displaced number state \cite{Glauber_1963}, and $\hat{\rho} = \ket{\Psi (t)} \bra{\Psi (t)}$ represents the density operator.\\
The presence of negativity in the Wigner function indicates that the associated state is non-classical. However, observing positive values throughout the Wigner function is not sufficient to conclude that the state is classical. Therefore, a state that exhibits a negative region in its phase-space distribution is inherently nonclassical~\cite{Kumar_2023}.

\begin{figure}[ht]
\centering
\includegraphics[width=1\linewidth]{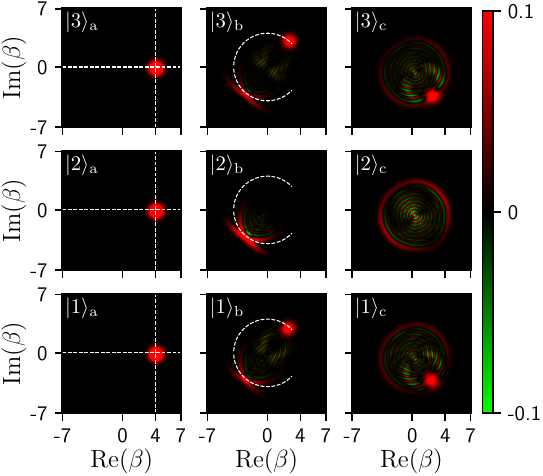}
\caption{The Wigner function  $W(\beta, \beta^*)$ corresponding to the same initial conditions given in the previous figures. In (a) for $ t = 0 $, in (b) for $t = 18 $, and in (c) for $ t = 45 $, corresponding to the different atomic levels: $\ket{3}$ upper level, $\ket{2}$ intermediate level, and $\ket{1}$ lower level, respectively.}
\label{fig5}
\end{figure}

In Fig.~\ref{fig5}, we show the Wigner function $W(\beta, \beta^*)$ corresponding to the same parameter values used in the previous figures, for different atomic conditions: when the atom starts in the upper level $\ket{3}$, the intermediate level $\ket{2}$, and the lower level $\ket{1}$. The subscripts a, b, and c represent the cases corresponding to $t=0$, $t=18$, and $t=45$, respectively. In Figs.~\ref{fig5}(a) ($t=0$), the Wigner function is well-localized and shifted 4 units to the left, as expected since the initial condition is a coherent state with $\alpha=4$. In Figs.~\ref{fig5}(b) ($t=18$) (see the red line in Fig. \ref{fig4}), a significant localized contribution can be observed in both the upper $\ket{3}$ and lower $\ket{1}$ levels, accompanied by a compressed region and small spots indicating the nonclassicality of the light. In the intermediate level $\ket{2}$, mostly classical light is observed. Finally, in Figs.~\ref{fig5}(c) ($t=45$), a greater contribution of non-classical light is observed, especially in the upper level $\ket{3}$, thus corroborating the results obtained from the analysis of Mandel's $\mathcal{Q} $ parameter.

To conclude this article, it is crucial to reiterate key findings regarding the classical and nonclassical behavior of quantum states. For instance, at $t=18$ for the lower level $\ket{1}$, Mandel's $\mathcal{Q}$ parameter suggests classical behavior. However, an analysis of the Wigner function reveals that nonclassical behavior is also present. It is essential to clarify that the negativity of the $\mathcal{Q}$ parameter is not a necessary criterion for distinguishing between classical and nonclassical regimes; it is merely a sufficient condition. In other words, while a negative $\mathcal{Q}$ parameter confirms nonclassical behavior, its positivity does not conclusively indicate classical behavior. On the other hand, the Wigner function provides a more definitive criterion. The presence of negative regions in the Wigner function unequivocally indicates that the associated state is nonclassical. Conversely, observing only positive values in the Wigner function does not guarantee that the state is classical. Hence, the negativity of the Wigner function is a necessary condition for nonclassicality. Therefore, for a quantum state to be considered nonclassical, it must exhibit negative regions in its phase-space distribution. A state with such negative regions is inherently nonclassical~\cite{Gerry_Book, Kumar_2023}.

\section{Conclusions}\label{Conclusiones}
In this study, we have explored the interaction of a ladder-type three-level atom ($\varXi$) confined within a lossless cavity containing a single-mode electromagnetic radiation field. Specifically, we assume that the field resonates with the transition frequency between the atom levels.

Based on these premises, we have developed criteria to detect quantum nonclassicality, exploring how atomic initial conditions impact these properties in the coupled light field. We observed that Mandel's $\mathcal{Q}$ parameter exhibits significant oscillatory behaviors over time and atomic initial states, reflecting varying degrees of nonclassicality in the system. Specifically, we noted higher nonclassicality when the atom starts in the upper state $\ket{3}$, while predominantly classical behavior was observed when the atom starts in the lower state $\ket{1}$.

Furthermore, we utilized the Wigner function to examine phase-space distributions of quantum states. The presence of negativity in the Wigner function was identified as a definitive indicator of nonclassicality, complementing the results obtained from Mandel's $\mathcal{Q}$ parameter. Visual representations of the Wigner function at different time points and atomic conditions provided visual confirmation of the observed nonclassical behavior in our study.

In summary, our study underscores the importance of atomic initial conditions in the manifestation of light nonclassicality in the ladder-type three-level  Jaynes-Cummings model ($\varXi$). The combination of Mandel's $\mathcal{Q}$ parameter and the Wigner function offers a powerful tool for characterizing nonclassical quantum states, highlighting the complexity and richness of quantum phenomena in light-matter systems.

\section*{Acknowledgments}
L. Hernández-Sánchez acknowledges the Instituto Nacional de Astrofísica, Óptica y Electrónica (INAOE) for the collaboration scholarship granted and the Consejo Nacional de Humanidades, Ciencias y Tecnologías (CONAHCYT) for the SNI III assistantship scholarship (No. CVU: 736710).

%

\end{document}